%% file: main_text.tex
\documentclass[preprint,12pt,nonatbib]{elsarticle}

\usepackage{amssymb}
\usepackage{amsmath}
\usepackage[utf8]{inputenc}
\DeclareUnicodeCharacter{202F}{ }
\usepackage{array}
\usepackage{tabularx}
\usepackage{lipsum}
\usepackage{xurl}
\usepackage{multirow}
\usepackage{booktabs} 
\usepackage{siunitx}  
\usepackage{comment}
\usepackage{placeins}
\usepackage{arydshln} 
\usepackage{hyperref}
\usepackage[backend=biber, sorting=none, url=false, doi=true]{biblatex}
\newcolumntype{Y}{>{\centering\arraybackslash}X}
\newcommand{\iiitgs}{I\textsuperscript{3}TGS}

\journal{Acta Materialia}

\begin{document}

\begin{frontmatter}



\title{High-throughput Down-selection of Fusion Candidate Materials} 

\affiliation[label1]{organization={MIT Plasma Science and Fusion Center (PSFC)},
             city={Cambridge},
             postcode={02139},
             state={MA},
             country={USA}}
\affiliation[label2]{organization={MIT Department of Materials Science \& Engineering},
             city={Cambridge},
             postcode={02139},
             state={MA},
             country={USA}}
\affiliation[label3]{organization={MIT Department of Mechanical Engineering},
             city={Cambridge},
             postcode={02139},
             state={MA},
             country={USA}}
\affiliation[label4]{organization={MIT Department of Nuclear Science \& Engineering},
             city={Cambridge},
             postcode={02139},
             state={MA},
             country={USA}}

\author[label1,label2]{Elena Botica-Artalejo}
\author[label3]{Josh Randolph}
\author[label1]{Kevin B. Woller}
\author[label1]{Gregory Wallace \corref{cor1}}
\ead{wallaceg@mit.edu}
\author[label1,label4]{Michael P. Short}

\cortext[cor1]{Corresponding author:}

\begin{abstract}
Developing radiation-resistant fusion reactor materials is constrained by the slow pace of conventional irradiation qualification methodologies, which usually require extensive one-at-a-time irradiation campaigns and post-irradiation characterization. At the current pace of development, we may not discover the materials necessary to deploy fusion power plants at desired rates. In this work, we present a high-throughput (HT) experimental framework for the accelerated down-selection of fusion-relevant, Cu-Cr-Ta RF antenna materials by integrating combinatorial synthesis, \textit{in situ} ion irradiation, and multi-property screening within a unified workflow. A 2D compositional Cu-Cr-Ta gradient wafer was fabricated by physical vapor deposition and characterized using \textit{in situ} ion irradiation transient grating spectroscopy (I$^3$TGS). Radiation-induced surface acoustic wave frequency fluctuations were used as a rapid proxy for primary defect accumulation, while thermal diffusivity measurements were obtained simultaneously during irradiation. Complementary HT characterization included EDS compositional mapping, four-point probe electrical resistivity measurements, and waste disposal rating calculations to evaluate long-term activation behavior. This integrated methodology enabled an acceleration with respect to conventional irradiation approaches by more than 100 times, from months to hours to analyze 78 new materials.

\end{abstract}

\begin{graphicalabstract}
\includegraphics[width=\textwidth]{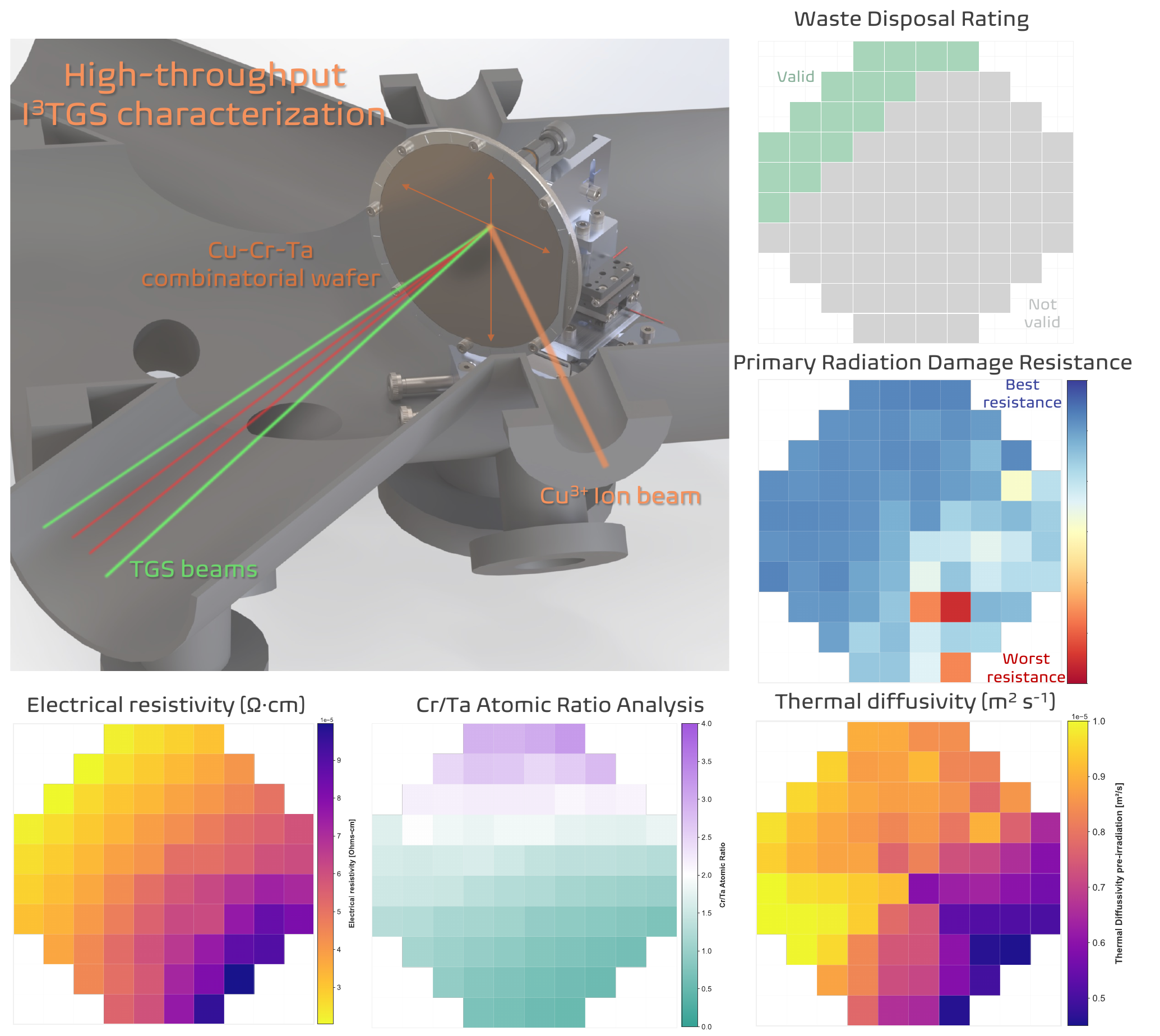}
\end{graphicalabstract}

\begin{highlights}
\item We have achieved high-throughput, \emph{in~situ} irradiation screening of fusion materials
\item Combinatorial synthesis and \emph{in~situ} ion irradiation hastens screening by 100x
\item Multimodal, relative changes in material properties identify the best composition
\item 78 Cu-Cr-Ta compositions were screened in 20 hours to optimize fusion RF antennas
\item Relying on instantaneous SAW speed drops during TGS was the key insight
\end{highlights}


\end{frontmatter}


\section{Introduction and Background}
One of the main challenges in developing practical fusion energy systems is the selection and sufficiently rapid qualification of materials capable of withstanding the conditions inside fusion reactors \cite{kurtz_chapter_2019, zinkle_motivation_2016}. Structural and plasma-facing components are simultaneously subjected to high heat fluxes, intense neutron irradiation, and severe thermo-mechanical stresses, and no existing material fully meets the requirements for long-term operation for many mission-critical components. Therefore, new alloys and material systems with enhanced radiation tolerance, high-temperature strength, and low activation are essential for economically viable fusion power plants \cite{was_materials_2019}. Traditional materials development and qualification processes are inherently time-consuming and resource intensive, usually involving the sequential fabrication, irradiation, and characterization of individual candidate materials, which can take weeks to months just to test a single alloy. Thus, performing such iterative testing for multiple material combinations can take months to years per material system, which is a significant bottleneck given the urgency of advancing fusion technology. 

To addresses the aforementioned time constraints, high-throughput (HT) technologies have grown, both experimental and computationally, as a transformative tool for accelerating discovery and down-selecting materials \cite{couet_integrated_2022,ludwig_discovery_2019}. HT approaches enable processes like parallel synthesis and evaluation of large libraries of materials across compositional spaces, reducing the time time required to identify promising candidate materials relative to traditional workflows \cite{shahzad_accelerating_2024}. Some of the most successful approaches in the HT experimentation are the use of automated characterization pipelines using computer vision analysis coupled with robotics \cite{siemenn_using_2024}, and the use of thin films to obtain dense combinatorial material libraries, due to the ease of fabrication \cite{wang_comprehensive_2023}. Additionally, some of these automated workflows are strengthened by the use of data-driven optimization, like AI/ML integration \cite{bassett_workflow_2023, huang_integration_2025, cheng_artificial_2026}.

The key insight in the successful implementation of HT materials down-selection technologies relies on asking the correct questions, in the correct order, to maximize information throughput in minimum time. Practically speaking, we must eschew the temptation to evaluate any absolute material properties in favor of relative ones in the first, most rapid steps of any HT framework. This is because there is no point making detailed measurements on a material which is decidedly not among the best few candidates, thus the incremental utility of ranking materials by \textit{relative} performance is far higher than absolute knowledge of any one at early stages. These HT frameworks often rely heavily on inference between properties of ultimate interest and those more easily measured, taking far less time and expense to achieve the same material ranking. Further optimizations can be made by relaxing bulk-like microstructural constraints, such as by using thick films in place of bulk samples, provided one can prove that changing the microstructure of all candidate materials uniformly does not alter their \textit{relative} ranking in performance. Therefore, by asking the fastest questions first and answering them with the coarsest, most relative information, one can most quickly arrive at the best materials worthy of more time- and resource-intensive measurements and/or computations.

Thin or thick film material generation of multiple materials at once is typically performed by vapor deposition (physical or chemical, PVD or CVD respectively), and less commonly using laser deposition techniques, like direct energy deposition (DED) or pulsed laser deposition (PLD)\cite{moorehead_high-throughput_2020, mcginn_thin-film_2019}. Using these approaches, it is possible to obtain continuous composition gradients along the substrate. However, in other cases, the composition is separated by a fiducial mask applied before the deposition techniques, or instead, using other approaches like ink‑jet/drop‑casting, or micro‑plates \cite{safran_review_2023, mao_high_2013}, to obtain a large number of discrete samples. To characterize these films, approaches like EDS, HT‑XRD, X-ray fluorescence (XRF), wavelength dispersive spectroscopy (WDS) or Rutherford backscattering spectroscopy (RBS) can be used to scan many positions on a film to map composition and phase across gradients or pad arrays \cite{ludwig_discovery_2019}. On the other hand, to explore microstructure and mechanical properties, other methods like electron backscatter diffraction (EBSD), nanoindentation, and micro-mechanical test arrays adapted to HT approaches \cite{wang_comprehensive_2023, safran_review_2023} are employed. Then, to assess functional properties, more specific techniques can be applied, for example 4-point probe for electrical resistivity, a profilometer to measure the film thickness or potential and Seebeck microprobe to measure the Seedbeck coefficient distribution, among others \cite{wambach_structural_2019}.

Nevertheless, one of the main challenges remaining in HT experiments using thin or thick films is the extrapolation of the properties to bulk microstructures \cite{vecchio_high-throughput_2021}. Thin and thick films both often exhibit microstructures, residual stresses, and grain structures that can differ significantly from those of bulk counterparts due to rapid deposition kinetics, substrate constraints (lattice mismatch, adhesion), and intrinsic dimensional effects \cite{wang_comprehensive_2023}. These differences can strongly influence \textit{absolute} mechanical, corrosion, and radiation‑response properties, making direct translation of film data to bulk behavior non‑trivial. Some studies have shown that while thin films may capture phase formation trends, their microstructural uniformity and fine grain sizes can lead to hardness and deformation responses that do not directly predict bulk mechanical performance \cite{wanni_assessing_2025}. Consequently, film-based results must be interpreted with caution and complemented by targeted bulk validation and a solid understanding of structure-property relationships to ensure that the observed properties can be safely extrapolated to bulk samples. However, because we seek relative rankings of material performance rather than absolute bulk properties, our constraints are significantly more relaxed than one may initially have thought.

Focusing on nuclear materials, recent advances in HT experiments have demonstrated integrated frameworks that couple rapid alloy synthesis, parallel irradiation, and modeling-informed down-selection. A representative example is the high-throughput synthesis of Mo-Nb-Ta-W refractory high-entropy alloys (HEAs) via \textit{in situ} alloying using additive manufacturing (AM), where arrays of discrete compositions were printed on a single build plate to efficiently explore a broad compositional space \cite{moorehead_high-throughput_2021}. When coupled with CALPHAD predictions, this approach established a robust platform for rapidly linking composition, phase stability, and baseline microstructure prior to irradiation. Complementing this accelerated synthesis, the development of a versatile, high-temperature, high-throughput ion irradiation system capable of exposing multiple samples simultaneously addressed a critical bottleneck in nuclear materials qualification by enabling controlled, parallel irradiation experiments under reactor-relevant conditions \cite{moorehead_development_2021}. The integration of these capabilities was subsequently demonstrated through HT ion irradiation of the additively manufactured compositionally complex alloys, where irradiation-induced microstructural evolution and hardening trends were mapped across composition space and interpreted in conjunction with thermodynamic modeling, shortening campaign time from months to weeks.

Even with this impressive achievement, we remain incredibly data-poor when it comes to knowledge of nuclear material performance. With dozens of potential parameters spanning spaces of composition, irradiation parameters, microstructure, and environmental compositions, one could not hope to leverage the upcoming power of artificial intelligence (AI) and machine learning (ML) to rapidly predict the best nuclear material for a given application. A cautionary tale can be found in the work of Jin et al. \cite{jin_computational_2019}, where the authors identified 27 independent parameters controlling radiation void swelling incubation doses in the roughly 370 catalogued experiments in human history. The entire idea of a nuclear materials laboratory must therefore be redefined in order to generate enough data from which AI models can learn and predict. By combining \textit{in situ} ion irradiation transient grating spectroscopy (I$^3$TGS) \cite{dennett_real-time_2019} with computationally informed combinatorial synthesis (CICS) \cite{botica-artalejo_real-time_2026}, we have shown the ability to hasten the evaluation of each nuclear material by 1,000,000x at the unit process. Accounting for material batch fabrication, accelerator pumping, heater stabilization, and ion source conditioning reduces this to factors of 100-1,000x, still approaching fast-enough.

In the present work, a novel methodology for the accelerated manufacture and down-selection of nuclear materials is realized and tested. PVD is employed to fabricate a combinatorial wafer containing a continuous compositional gradient within the Cu-Cr-Ta system. This was chosen as a low-activation replacement of the Cu-Cr-Nb systems proposed in previous heating and current drive studies which do not properly consider radiation concerns. This alloy has been previously selected as a promising material system for radio-frequency (RF) antenna systems in fusion reactors \cite{botica-artalejo_rapid_2026}. Later, this wafer is subsequently subjected to \emph{in situ} self-ion irradiation coupled with transient grating spectroscopy (TGS). Using this technique, radiation damage resistance is quantified in the initial stages of irradiation when point defect saturation is reached, thanks to the established correlation between initial SAW frequency decay and void swelling \cite{botica-artalejo_rapid_2026}, therefore a spatially resolved map of primary radiation damage resistance is obtained across the compositional space. By integrating these results with thermal diffusivity, electrical resistivity measurements, and stoichiometric analysis to assess the expected formation of secondary phases, a data-driven down-selection of the most promising alloy composition is performed. The present methodology is used within the Cu-Cr-Ta system as showcase, to identify an optimized candidate material selection for RF antenna applications in fusion reactors.

\section{Methods}
\subsection{High-throughput material generation}
\begin{figure}
    \centering
    \includegraphics[width=\textwidth]{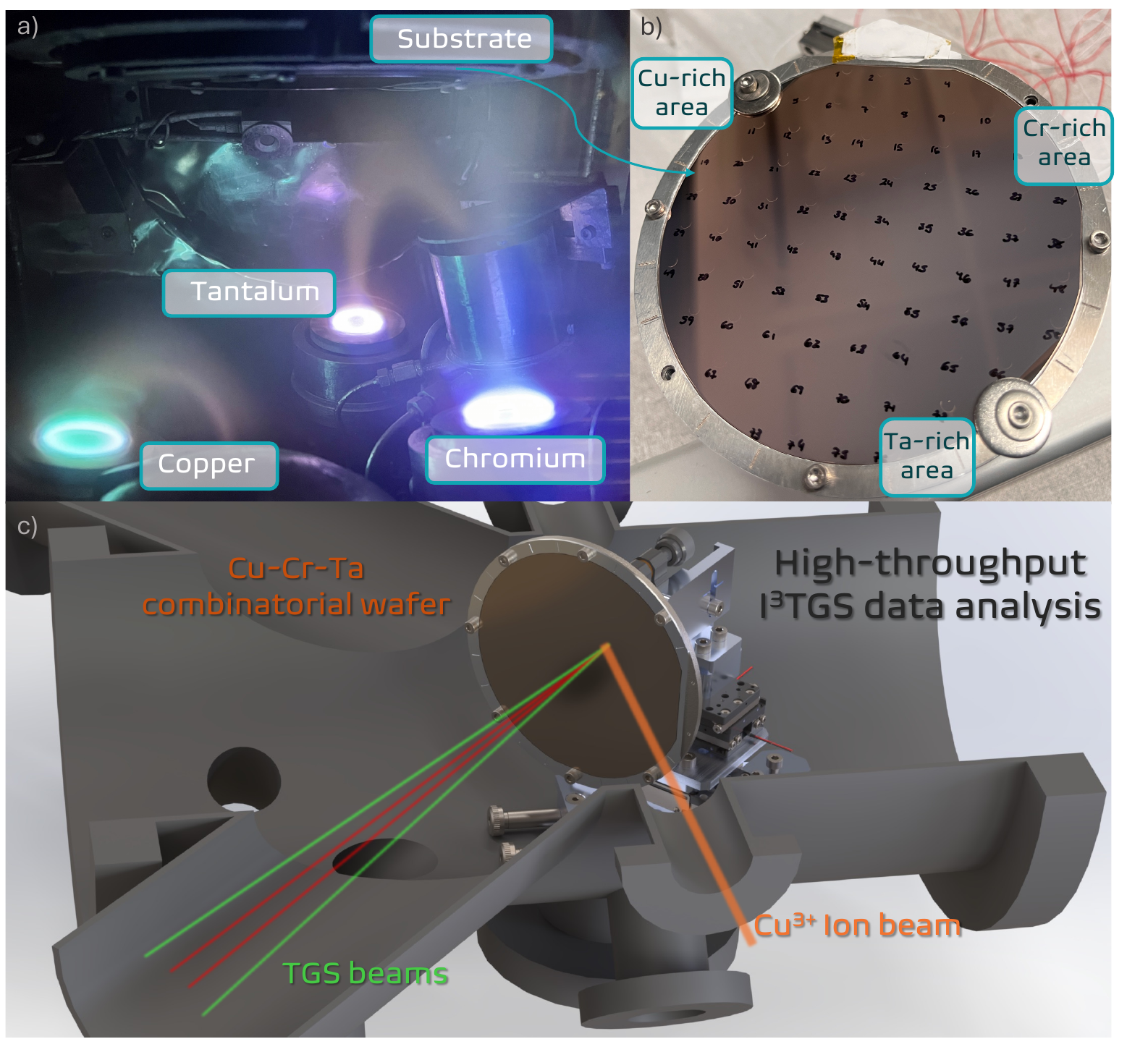}
    \caption{\small a) Combinatorial PVD deposition of Cu-Cr-Ta alloys. b) Deposited 4-inch wafer with fiducial marks and numbers for spot identification. c) Render of the $\mathrm{I^3TGS}$ experiment using the new multi-axial stage.}
    \label{fig:PVD_I3TGS_lay}
\end{figure}

Cu-Cr-Ta combinatorial thick films were created using PVD in a Kurt J. Lesker PVD 75 PRO Line. It allows simultaneous deposition of up to five different elements (our system's configuration allows for three), in which each sputtering target is independently controlled. Spatial variations are laid down by each sputtering target, utilizing different target sizes, powers, off-axis positions, and angles to control relative elemental deposition rates and gradient slopes. This creates a sample with compositional gradients in two dimensions across the substrate (see Figure~\ref{fig:PVD_I3TGS_lay}), in our cases building a ternary phase diagram on a wafer. The specific parameters used for PVD deposition in this study can be seen in Table~\ref{tab:deposition_parameters}. To have a better film quality during the magnetron sputtering process, the base pressure was in the 10$^{-6}$ range and the Ar pressure during deposition was kept at $5.9 \times 10^{-4}$ Torr. The sputtering power applied to each target was tuned to control the relative concentration of each constituent element. High-purity targets from Kurt J. Lesker were used: Cu (99.99\%), Cr (99.95\%), and Ta (99.95\%). Certifications for target impurities are provided in the associated data repository. Before deposition, a Ti adhesion layer (99.995\%) was deposited to improve film adhesion to the substrate and serve as an oxygen getter, thereby minimizing contamination. The coatings were deposited onto 1-inch p-type (B-doped) semi-prime silicon wafers with a surface roughness of $R_a < 0.5\mu m$.
\begin{table}[htbp]
\centering
\caption{Deposition parameters for Cu-Cr-Ta thick film.}
\label{tab:deposition_parameters}
\small

\begin{tabularx}{\textwidth}{>{\raggedright\arraybackslash}X c}
\toprule
\textbf{Parameter} & \textbf{Value} \\
\midrule
Vacuum level prior to deposition (Torr) & \(6.8 \times 10^{-6}\) \\
Chamber pressure during deposition (Torr) & \(5.9 \times 10^{-4}\) \\
Sputtering time (min) & 150 \\
Gas flow (sccm) & 17 \\
Gas species & Argon (99.999\%) \\
Temperature (\(^\circ\)C) & 20--25 \\
Cu target power \& size (Watts/inch/type) & 148/3/DC \\
Cr target power \& size (Watts/inch/type) & 40/2/RF \\
Ta target power \& size (Watts/inch/type) & 18/2/DC \\
Ti target power \& size (Watts/inch/type) & 60/2/DC \\
Ti sputtering time (min) & 60 \\
\bottomrule
\end{tabularx}
\end{table}

\subsection{Microscopy characterization}
The elemental composition of each wafer was characterized by energy-dispersive X-ray spectroscopy (EDS) using a Zeiss Gemini 450 scanning electron microscope (SEM). An accelerating voltage of 25 kV was employed to effectively excite the characteristic X-ray lines while probing a sufficient material depth. The working distance was fixed at 8.5 mm to maximize X-ray collection.

\subsection{In situ ion irradiation \texorpdfstring{\iiitgs}{I3TGS}}
TGS is a non-contact, laser-based method used to quantitatively measure a material's elastic properties and thermal diffusivity at controlled depths \cite{hofmann_transient_2019}. This technique provides localized measurements of the thermo-elastic response with a spatial resolution determined by the focal width of the probe laser. We specifically use these thermoelastic property changes as a function of time and composition as our inference model for ultra-rapid radiation damage resistance assessment \cite{botica-artalejo_rapid_2026}. Firstly, a pump laser ($1kHz$, $532nm$, $\sim 500ps$ pulses) excites the material surface, generating a grating pattern with an on-sample power of $7.4mW$. The grating spacing determines the maximum depth for probing thermo-elastic properties. Then, a continuous-wave probe laser (577 nm) monitors the surface excitation via first-order diffraction, with an intensity of approximately $127kW m^{-2}$. The probe beam measures the surface acoustic wave (SAW) frequency, which relates to elastic properties, and the signal decay provides the thermal diffusivity. 

A 0.2 neutral density filter was employed to reduce the on-sample pump power to $5.89 mW$. A di-homodyne phase collection geometry was used \cite{dennett_thermal_2018}, employing two phase-locked probe beams ($\phi = \pm \tfrac{\pi}{2}$) to stabilize the signal against phase fluctuations. The current TGS experimental setup is detailed in \cite{botica-artalejo_rapid_2026}. The probing depth was adjusted by changing the grating spacing, to match the irradiated region of the material. The grating space $\Lambda$ has a nominal sensitivity of $\lambda/\pi$ for thermal properties \cite{kading_transient_1995} and $\lambda/2$ for SAW signals \cite{royer_rayleigh_1984} -- in reality signals are returned with roughly exponentially diminishing strength from each successive slice of material. The distances stated here correspond to the \emph{e}-folding depths of signal return. Details of all relevant TGS system parameters employed can be seen in Table~\ref{tab:tgs_param} from Supplementary material. Time resolution was optimized and settled on 2000 traces per 10 seconds, which improved temporal resolution without noticeably compromising data quality. A custom MATLAB code processed raw TGS data to extract SAW frequency and thermal diffusivity \cite{short_lab_github_2025}. Both raw and processed data are available in this article’s data repository \cite{botica-artalejo_github_2026}.

The TGS system was mounted pointing to the end-of-beamline chamber viewport, enabling \emph{in situ} monitoring during irradiation. Samples were rotated 45$^\circ$ relative to the ion beam to maintain perpendicular alignment with the TGS system, required for di-homodyne signal formation. Irradiations were performed at the Cambridge Laboratory for Accelerator Surface Science (CLASS) at the MIT Plasma Science and Fusion Center, using a Tandetron tandem ion accelerator at 1.65 MV. Experiments employed Cu$^{3+}$ ions at 6.6 MeV and a flux rate of $9.4\times 10^{15} ions/m^2\cdot s$. Beam current was monitored with a picoammeter connected to the sample holder, isolated electrically from the rest of the beamline. A 2 $mm$ aperture focused the beam, producing an ellipsoidal spot of 4.44 $mm^2$ on the angled sample. Prior to experiments, beam alignment with the TGS spot was verified using a silver-activated zinc sulfide coating on a ceramic substrate 1 $mm$ thick laid at the side of the wafer. Ion penetration depth and damage profiles were simulated using SRIM \cite{ziegler_srim_2010}, following the procedure indicated in \cite{stoller_use_2013} with modified displacement energies of 30 eV for Cu, 40 eV for Cr and 90 eV for Ta, according to \cite{astm_astm_2023}, obtaining a penetration depth of $\sim 2\mu m$. Material densities were automatically calculated by SRIM and utilized as-is.

In order to characterize the radiation damage resistance across the entire 4-inch diameter surface of the wafer, a dedicated sample stage was designed to enable spatially resolved measurements while maintaining the optical and irradiation constraints of the experimental setup. In the I$^3$TGS configuration, the TGS laser system is fixed at a defined focal distance, and the ion beam is aligned to impinge on the same interaction region probed by the TGS measurement. Consequently, full mobility of the wafer is required to ensure that any position on its surface can be accurately placed at the common focal and irradiation point. To satisfy these requirements, a five-axis, vacuum-compatible stage was designed and fabricated, see Figure \ref{fig:PVD_I3TGS_lay}.

\subsection{Electrical resistivity}
To determine the electrical resistivity of each composition within the combinatorial thick film, four-point probe measurements were performed using a Pro4 system. The instrument operates in the in-line four-point probe configuration in accordance with ASTM F84 standards. Four-point probe returns the sheet resistance, $R_s$, which relates with the electrical resistivity ($\rho$) through the thickness ($t$) of the measured material as follows:

\begin{equation}
\label{eq:sheet_res}
    R_s=\frac{\rho}{t}
\end{equation}

Resistivity mapping was carried out by rastering the wafer beneath the probe to systematically sample different compositions across the film. Measurements were collected at $1 cm$ intervals to ensure adequate spatial resolution while covering the full compositional gradient.

\subsection{CALPHAD calculations}
CALPHAD calculations were carried out using Thermo-Calc 2024b in combination with the TCHEA7 and TCHEA8 thermodynamic databases for high-entropy alloys. Thermal and electrical conductivity predictions were performed across compositional ranges spanning from pure Cu to pure Cr$_2$X compounds, where X represents elements known to form C15 Laves phases analogous to Nb, including Zr, Ta, and Hf \cite{kumar_structural_2000}. In addition, equilibrium phase stability calculations were conducted for the most promising alloy candidate identified for RF antenna applications in order to assess the thermal stability of the system over the anticipated operating temperature range.

\section{Results}
\subsection{2D array compositional mapping}
The chemical composition and spatial elemental distribution across the combinatorial wafer were characterized by energy-dispersive X-ray spectroscopy (EDS) and scanning electron microscopy (SEM). Prior to deposition, fiducial marks separated by 1 cm were engraved onto the substrate to provide positional references during compositional mapping and ensure consistent identification of each measurement location. The reported composition at each position corresponds to the average of three area-integrated EDS measurements, thereby minimizing local statistical fluctuations associated with single-point analyses.

\begin{figure}
\centering
\includegraphics[width=\textwidth]{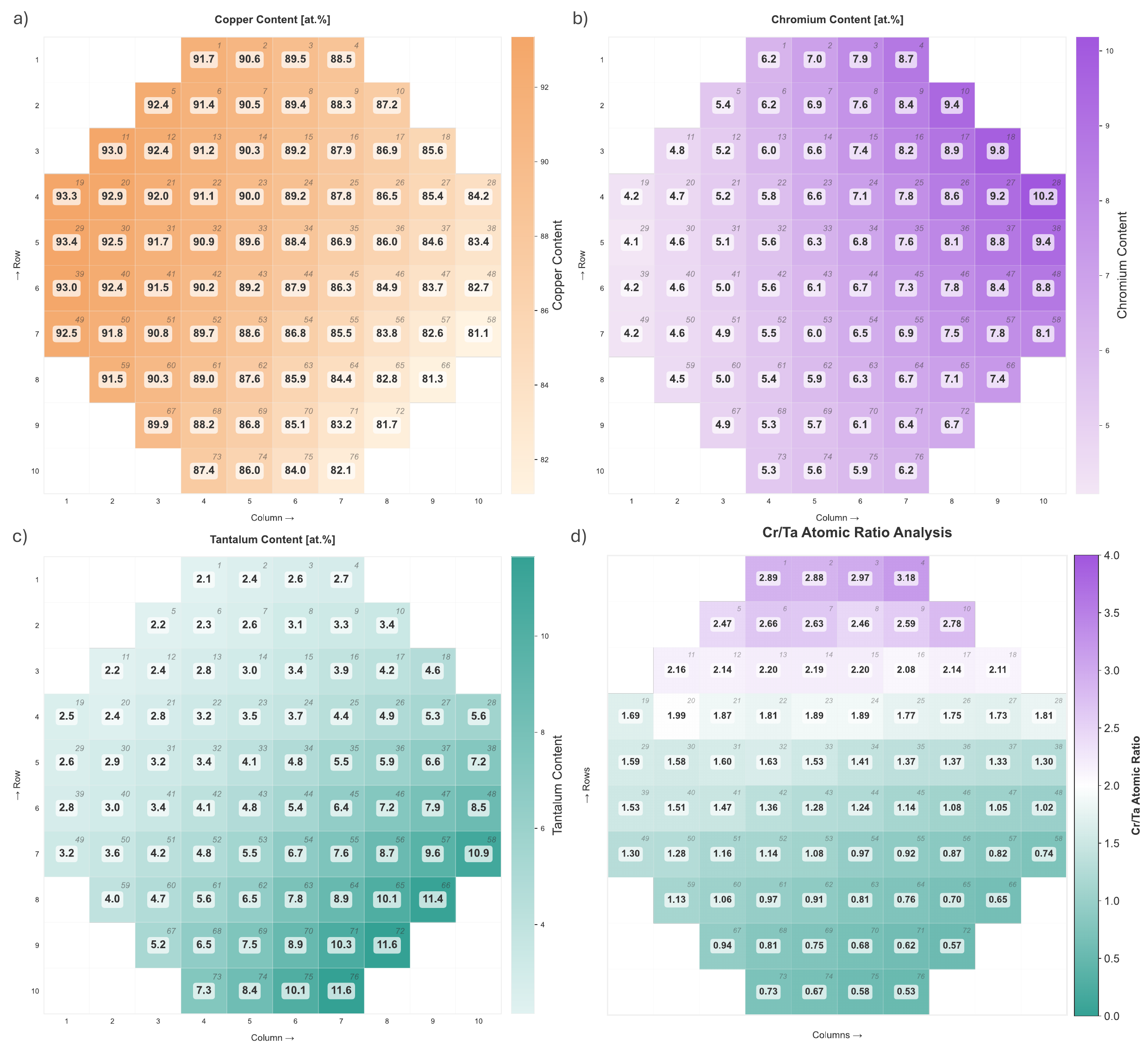}
\caption{EDS compositional mapping of the Cu-Cr-Ta combinatorial wafer. Panels a)-c) show the spatial distribution of Cu, Cr, and Ta, respectively. Panel d) presents the Cr/Ta ratio, where lighter regions correspond to compositions approaching Cr/Ta = 2, favorable for Cr$_2$Ta precipitate formation.}
\label{fig:chem_comp}
\end{figure}

Figure \ref{fig:chem_comp} shows the spatial distribution of Cu, Cr, and Ta across the deposited wafer, together with the corresponding Cr/Ta ratio. This ratio is of particular interest because compositions approaching Cr/Ta $\approx 2$ favor the formation of Cr$_2$Ta precipitates, which can promote the development of C15 Laves phases \cite{kumar_structural_2000}. Such precipitates are expected to enhance the high-temperature mechanical strength of the alloy system, analogous to the strengthening mechanisms reported in GRCop-84 alloys \cite{ellis_grcop-84_2005}.

\subsection{High-throughput radiation damage characterization using \texorpdfstring{\iiitgs}{I3TGS}}
The radiation damage response of the Cu-Cr-Ta alloy system was evaluated using \emph{in~situ} ion irradiation transient grating spectroscopy (I$^3$TGS)~\cite{dennett_real-time_2019}. In addition to enabling real-time assessment of irradiation-induced elastic property changes, the measurements simultaneously provided thermal diffusivity values for each composition. Thermal transport is a critical parameter for RF antenna applications, where efficient dissipation of plasma-induced heat loads is required to maintain structural integrity during operation.

To quantify radiation damage resistance, an automated Python-based analysis routine was developed to extract the maximum decrease in SAW frequency during irradiation. The reduction in SAW frequency corresponds to elastic softening associated with point defect accumulation and reaches a minimum once defect saturation is achieved \cite{botica-artalejo_rapid_2026}. Consequently, the relative SAW frequency drop at saturation was used as an indirect metric of the retained defect population (see Figure \ref{fig:I3TGS_beam} of Supplementary material) and, therefore, as a proxy for primary radiation damage resistance. Smaller frequency shifts indicate reduced defect production, leading to enhanced resistance to irradiation-induced damage accumulation. This can be explained as lower vacancy concentrations at point defect saturation level are expected to reduce the probability of subsequent defect clustering and void swelling under continued irradiation, thereby mitigating long-term microstructural degradation and associated property deterioration. The automated analysis workflow used for signal extraction and processing is available in the associated repository \cite{botica-artalejo_github_2026}. 

\begin{figure}
\centering
\includegraphics[width=\textwidth]{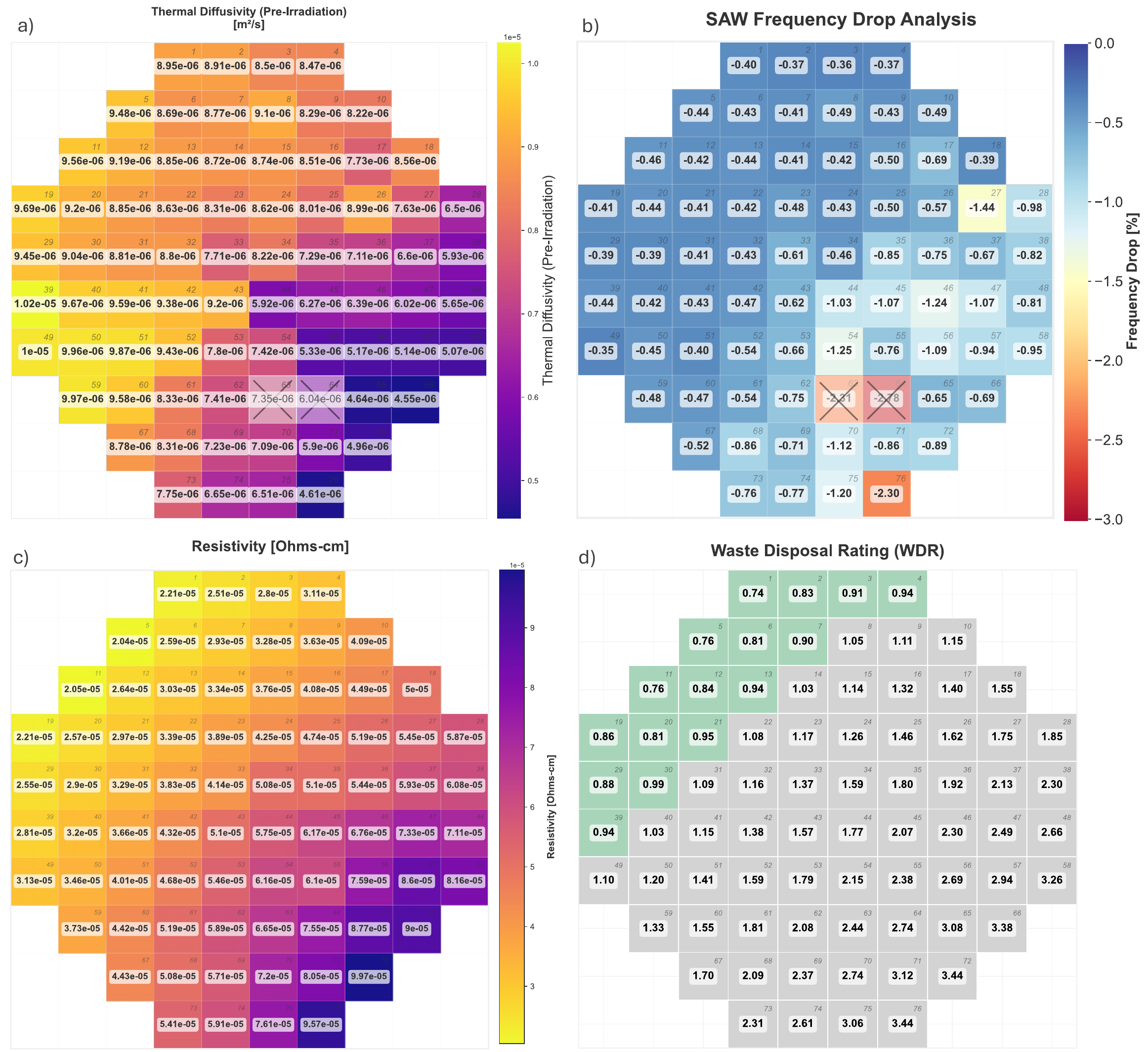}
\caption{High-throughput I$^3$TGS characterization of the Cu-Cr-Ta combinatorial wafer. Panels a) shows the pre-irradiation thermal diffusivity values, respectively. Panel b) presents the calculated SAW frequency drop induced by ion irradiation. Measurement locations 63 and 64 were excluded due to surface contamination introduced during sample cleaning. Panel c) shows the electrical resistivity distribution and panel d) represents the Waste disposal rating (WDR) map. Compositions classified as low-level waste (LLW) are highlighted in green, whereas high-level waste (HLW) compositions are shown in gray. The orientation matches that of the Figure \ref{fig:chem_comp}, allowing each value to be directly associated with the corresponding composition}
\label{fig:HT_prop}
\end{figure}

Figures \ref{fig:HT_prop}.a and \ref{fig:HT_prop}.b summarize the spatial distribution of the measured thermal diffusivity pre-irradiation and the irradiation-induced SAW frequency drop across the wafer, respectively. The data are displayed in the original wafer geometry to facilitate direct comparison with the compositional distributions shown in Figure \ref{fig:chem_comp}. Measurement locations 63 and 64 were excluded from further analysis due to surface contamination introduced during the cleaning procedure prior to the I$^3$TGS experiment. Specifically, the use of a compressed-gas duster resulted in localized deposition of chemical residues that altered both the surface composition and optical reflectivity, compromising the reliability of the TGS measurements. The affected region is visible in Figure \ref{fig:HT_prop}.

\subsection{Electrical resistivity measurements}
Electrical conductivity is another key performance requirement for RF antenna applications. Therefore, four-point probe measurements were performed across the combinatorial wafer at the same locations used for the I$^3$TGS characterization. 

The four-point probe technique provides the sheet resistance of the deposited films. Accordingly, the electrical resistivity was calculated by combining the measured sheet resistance with the local film thickness according to Equation \ref{eq:sheet_res}. Based on previous depositions performed under identical conditions, the film thickness was measured to be approximately $5.3~\mu$m at the wafer center and $3.8~\mu$m near the edge. A linear interpolation was used to estimate the thickness at intermediate radial positions. Figure \ref{fig:HT_prop}.c presents the resulting electrical resistivity distributions in the original wafer geometry.

\subsection{Waste disposal rating calculations}
The waste disposal rating (WDR) quantifies the long-term activation behavior of materials exposed to neutron irradiation and provides a criterion for classifying materials as low-level waste (LLW) or high-level waste (HLW). Materials with WDR values below 1 are classified as LLW, whereas values above 1 correspond to HLW \cite{el-guebaly_managing_2023}.

Current U.S. Nuclear Regulatory Commission (NRC) regulations under 10 CFR 61 \cite{nuclear_regulatory_commission_part_2025} consider only a limited set of radionuclides and actinides \cite{klueh_impurity_2000, el-guebaly_design_2017}, which does not fully capture the activation behavior associated with the 14 MeV neutron spectrum characteristic of fusion environments. To address this limitation, the activation limits proposed by Fetter et al. \cite{fetter_long-term_1990} were adopted in the present work. These limits were derived considering neutron fluxes representative of the first wall region of a lithium-cooled fusion blanket system with a vanadium-alloy vacuum vessel and an accumulated neutron fluence of $20~\mathrm{MW\cdot y/m^2}$. Because the exact coolant and vacuum vessel materials for the proposed antenna system remain undefined, the upper-bound activation limits proposed by Fetter et al. were used to provide a more flexible while conservative assessment. Figure \ref{fig:HT_prop}.d presents the calculated WDR map for the deposited compositions, where LLW and HLW compositions are distinguished accordingly by green and grey colors, respectively.  The regulation of radioactive materials from fusion reactors is an evolving field and WDR will need to be reevaluated as regulatory standards and fusion reactor operating scenarios mature.  The WDR values for the Cu-Cr-Ta system considered here are a factor of $\sim 10^3$ smaller than for the GRCop Cu-Cr-Nb family of alloys.

\subsection{CALPHAD estimation of candidate alloy thermal stability}
CALPHAD calculations were performed for the highest-ranked alloy candidate identified through the multi-criteria screening methodology in order to evaluate its phase stability within the anticipated RF antenna operating temperature range ($\sim700^\circ$C). The calculations were carried out using Thermo-Calc 2025b in combination with the TCHEA8 database, as the available Cu-specific database did not include Ta-containing systems. Figure \ref{fig:TC_phase} of Supplementary material shows the predicted phase fractions as a function of temperature, showing that the alloy system has solid stable phases at $\sim700^\circ$C.

\section{Discussion}
\subsection{Relevance and limitations of transport property measurements in combinatorial thick films}
The transport properties reported in this work were obtained from PVD-grown thick films rather than fully dense bulk materials. Consequently, the measured electrical and thermal transport values are influenced by microstructural features characteristic of deposited films, including refined grain sizes, elevated defect densities, and residual stresses introduced during growth \cite{thornton_influence_1975}. These features increase electron and phonon scattering, generally resulting in lower electrical conductivity and thermal diffusivity than would be expected in bulk counterparts.
However, the objective of the present study is not the absolute quantification of engineering-ready bulk properties, but rather the rapid comparative screening of a broad compositional space under controlled and reproducible conditions. Since all compositions were synthesized using identical deposition parameters and characterized using the same experimental workflow, the measured trends remain internally consistent and therefore highly suitable for comparative down-selection. In this context, the combinatorial dataset should be interpreted as a relative performance map that enables rapid identification of promising compositional regions prior to bulk-scale validation.

An additional consideration concerns the relevance of DC electrical conductivity measurements for RF antenna applications. Ideally, conductivity should be evaluated directly at RF or microwave operating frequencies. Nevertheless, high-throughput RF conductivity measurements remain experimentally challenging due to the lack of standardized rapid-screening methodologies compatible with combinatorial workflows. Techniques such as scanning microwave microscopy (SMM) \cite{tami_scanning_2025} and microwave impedance microscopy (MIM) \cite{barber_microwave_2021} provide localized conductivity measurements at nanometer length scales, but such measurements are strongly affected by local microstructural heterogeneity and are therefore not necessarily representative of the macroscopic electrical behavior relevant for RF antenna operation.

In contrast, four-point probe measurements provide an area-averaged assessment of conductivity and are readily scalable to high-throughput characterization campaigns. Furthermore, according to the classical Drude formalism \cite{smith_classical_2001}, where
\begin{equation}
    \sigma (\omega) = \frac{\sigma _0}{1+\omega^2 \tau^2}
    \label{eq:drude}
\end{equation}
where $\sigma_0$ is the DC conductivity, $\omega$ is the angular frequency and $\tau$ is the electron relaxation time, for RF antennas operating in the range of $10^{6}-10^{10}Hz$, the factor $\omega \tau <<1$ implies that equation \ref{eq:drude} can be reduced to $\sigma (\omega) \approx \sigma _0$, as the electric field changes slowly compared to the electron scattering events, which are able to respond instantaneously to fluctuations in the field. Under this approximation, DC conductivity provides a reasonable first-order estimate of RF conductivity for comparative screening purposes. Previous studies by Thorp \cite{thorp_rf_1954} further demonstrated that discrepancies between RF and DC conductivity are strongly influenced by surface roughness, residual stress, and abnormal surface layers, effects that can be mitigated through post-processing treatments such as annealing and surface polishing.
Taken together, these considerations support the use of DC transport measurements within the present high-throughput framework, particularly for rapid ranking and comparative evaluation of candidate compositions, and specifically noting the smooth and uniform nature of the thick deposited films used herein.

\subsection{High-throughput fusion materials down-selection framework}
A central objective of this work was the development of an accelerated experimental workflow capable of rapidly identifying promising fusion-relevant structural and functional materials. Conventional qualification routes for irradiation-resistant materials typically rely on sequential bulk alloy fabrication, long-duration irradiation campaigns, TEM-based defect characterization, and iterative optimization cycles, resulting in development timelines extending over months or years for a single alloy family. The methodology presented here substantially compresses this process by integrating combinatorial synthesis, \emph{in~situ} irradiation diagnostics, and multi-property screening into a unified experimental platform. In other words, focusing on qualification first rather than last is the field-specific definition of putting the cart before the horse.

As a proof of concept, a combinatorial Cu-Cr-Ta wafer spanning a broad compositional range [Cu (81-93 at.\%), Cr (4-10 at.\%), Ta (2-11 at.\%)] was fabricated by thick film PVD and characterized using a newly developed five-axis vacuum-compatible stage integrated into the I$^3$TGS system. This configuration enabled spatially resolved irradiation experiments to be performed sequentially across the entire wafer without venting the chamber or re-aligning the ion beam between measurements. As a result, 76 distinct alloy compositions were evaluated for relative RF antenna suitability under identical irradiation conditions within a single experimental campaign, reducing screening times by 100 times (from months to hours).

This capability represents a substantial increase in experimental throughput compared with conventional irradiation-based materials screening methodologies, provided relative ranking of materials is the objective. Importantly, the approach simultaneously provides multiple performance metrics, including irradiation-induced elastic softening which is related to primary radiation damage resistance, thermal diffusivity, electrical resistivity, compositional mapping, and activation behavior. The resulting dataset constitutes a multidimensional property library that can be directly interrogated to identify optimal compositional regions. Although a gap remains between film characterization and bulk engineering performance, the methodology is specifically designed to capture relative compositional trends rather than absolute engineering values during the initial discovery stage. The validity of this approach is supported by several considerations.

First, during irradiation experiments, the characteristic grain dimensions remained sufficiently larger than the estimated collision cascade diameter ($\sim5$ nm \cite{was_fundamentals_2017}), ensuring that primary defect generation processes were not dominated by grain boundary interactions. Consequently, the measured SAW frequency evolution primarily reflects intrinsic defect accumulation behavior associated with each composition.
Second, while electrical and thermal transport values are affected by film microstructure, all compositions were synthesized and characterized under comparable conditions. Therefore, relative variations across the compositional spread remain representative of intrinsic compositional effects and provide a reliable basis for comparative screening.

More broadly, the present workflow demonstrates how combinatorial materials science can be integrated with irradiation-aware diagnostics to accelerate fusion materials development. Rather than replacing conventional bulk qualification, this methodology functions as a rapid front-end discovery and prioritization platform that drastically reduces the number of candidate alloys requiring costly large-scale validation.

\subsection{Multi-criteria optimization of Cu-Cr-Ta alloys for RF antenna applications}
The Cu-Cr-Ta alloy system investigated in this work was designed to simultaneously satisfy several competing requirements associated with RF antenna operation in fusion reactors, including low neutron activation, high electrical and thermal conductivity, radiation damage resistance, and adequate high-temperature mechanical strength.

To address these competing constraints, a structured multi-criteria down-selection methodology was implemented. The first selection criterion was based on the waste disposal rating (WDR). Only compositions satisfying WDR $<1$ were retained, ensuring classification as low-level waste (LLW) materials and compatibility with reduced long-term activation requirements for fusion systems. A second filtering step targeted compositions with Cr/Ta ratios approaching $2\pm0.5$, corresponding to the compositional regime favorable for Cr$_2$Ta precipitate formation. This criterion was motivated by the expected development of C15 Laves secondary phases, which provide precipitation strengthening while minimizing solute content within the Cu-rich matrix. Such partitioning is advantageous because it enables simultaneous enhancement of mechanical strength and retention of thermal and electrical conductivity.

Following compositional pre-selection, the remaining candidates were ranked using an integrated property-difference methodology incorporating: (i) Cr/Ta ratio, (ii) irradiation-induced SAW frequency drop, (iii) thermal diffusivity, and (iv) electrical resistivity. Rather than applying arbitrary weighting factors, the ranking procedure quantified the deviation of each composition from the best-performing value for each property, enabling objective comparison across the compositional space (see Table \ref{tab:measured_ranking}). This integrated evaluation identified spot 11, corresponding to Cu-4.8 at.\%Cr-2.2 at.\%Ta, as the most promising candidate within the investigated alloy system. This composition exhibited an optimal balance between radiation damage resistance, thermal transport, electrical conductivity, and phase-selection criteria. At this point it is relevant to mention, that once the reactor design is completely optimized, the weight on each of the properties for the RF antennas can be changed to obtain a more adequate material system that prioritize specific features.

\begin{table}
\centering
\caption{Measured values and ranking for filtered spots, with "Cumulative" referring to a simple sum of property difference rankings, i.e. with equal weights. A more physically useful figure of merit (FOM) could be easily constructed to better weight property differences depending on the desired application.}
\label{tab:measured_ranking}
\resizebox{\linewidth}{!}{
\begin{tabular}{c|cccc|cccc|c}
\hline
\multirow{2}{*}{\textbf{Spot}} 
& \multicolumn{4}{c|}{\textbf{Measured value}} 
& \multicolumn{4}{c|}{\textbf{Property Difference Ranking}} 
& \multirow{2}{*}{\textbf{Cumulative}} \\
& Cr/Ta 
& $f_{SAW}$ drop [\%] 
& $\alpha _{pre}$ [m$^2$s$^{-1}$]
& Resistivity [$\Omega \cdot$cm] 
& 2-(Cr/Ta)
& $f_{SAW}$ drop 
& $\alpha _{pre}$
& Resistivity 
& \\
\hline
11 & 2.16 & 0.46 & $9.56\times 10^{-6}$ & $2.05\times 10^{-5}$ & 0.16 & 0.07 & 0.06 & 0.01 & 0.30 \\
19 & 1.69 & 0.41 & $9.96\times 10^{-6}$ & $2.21\times 10^{-5}$ & 0.31 & 0.02 & 0.02 & 0.17 & 0.52 \\
5  & 2.47 & 0.44 & $9.48\times 10^{-6}$ & $2.04\times 10^{-5}$ & 0.47 & 0.05 & 0.07 & 0.00 & 0.59 \\
20 & 1.99 & 0.44 & $9.20\times 10^{-6}$ & $2.57\times 10^{-5}$ & 0.01 & 0.05 & 0.10 & 0.53 & 0.69 \\
12 & 2.14 & 0.42 & $9.19\times 10^{-6}$ & 2$.64\times 10^{-5}$ & 0.14 & 0.03 & 0.10 & 0.60 & 0.87 \\
29 & 1.59 & 0.39 & $9.45\times 10^{-6}$ & $2.55\times 10^{-5}$ & 0.41 & 0.00 & 0.08 & 0.51 & 0.99 \\
21 & 1.87 & 0.41 & $8.85\times 10^{-6}$ & $2.97\times 10^{-5}$ & 0.13 & 0.02 & 0.14 & 0.93 & 1.21 \\
39 & 1.53 & 0.44 & $1.02\times 10^{-5}$ & $2.81\times 10^{-5}$ & 0.47 & 0.05 & 0.00 & 0.77 & 1.29 \\
13 & 2.20 & 0.44 & $8.85\times 10^{-6}$ & $3.03\times 10^{-5}$ & 0.20 & 0.05 & 0.14 & 0.99 & 1.37 \\
30 & 1.58 & 0.39 & $9.04\times 10^{-6}$ & $2.90\times 10^{-5}$ & 0.42 & 0.00 & 0.12 & 0.86 & 1.40 \\
\hline
\end{tabular}}
\end{table}

Importantly, the ability to identify this candidate emerged not from optimization of a single property, but from the simultaneous integration of multiple performance metrics within a high-throughput workflow. This highlights one of the principal strengths of the present methodology: enabling rapid navigation of complex multi-dimensional materials design spaces relevant to fusion applications.

Finally, CALPHAD phase stability calculations were performed for the selected alloy composition to evaluate its expected thermal stability under anticipated operating conditions. The calculations indicate that the alloy maintains phase stability up to approximately $1084^\circ$C, at which point the FCC Cu-rich phase melts while Cr$_2$Ta precipitates remain stable. This predicted stability exceeds the anticipated RF antenna operating temperature ($\sim700^\circ$C), supporting the suitability of the selected composition for high-temperature fusion applications.

\subsection{Implications for Artificial Intelligence Materials Discovery}
The field of artificial intelligence (AI) has exploded onto the scene in terms of materials discovery \cite{cheng_artificial_2026}. However, the potential usage of AI for \emph{practically useful} materials discovery depends strongly on the quality of the data used to train its underlying models, and especially the quantity of data in addition to its quality \cite{papadimitriou_ai_2024}. Efforts like the Material Project have identified many new crystal structures and compounds from first-principles calculations \cite{merchant_scaling_2023}, while a combination of existing data and first-principles calculations have identified highly efficient solar cell materials \cite{wu_inverse_2024}. However, our paper exemplifies a very different class of materials discovery problems, endemic not only to radiation materials science and nuclear materials, but also to physical metallurgy, functional ceramics, and similar fields. The issues revolve around two interrelated phenomena - data scarcity, and data quality \cite{reeves-mclaren_data_2026}. Scarcity is on full display in the literature, for example Jin et al. had previously attempted to discover hidden trends in radiation void swelling in metals~\cite{jin_computational_2019}. However, the researchers identified 27 almost linearly independent parameters among a dataset of fewer than 400 experiments yet conducted in human history, precluding the deep learning of absolutely anything.

The high-throughput combinatorial materials discovery framework described in this manuscript promises to turn this narrative on its head, by creating voluminous data of sufficient, but not necessarily optimal, quality from which models can learn. By speeding up materials characterization by factors of 100 in this study, or improving data throughput by factors of 1,000,000 through a combination of our previous studies~\cite{dennett_real-time_2019,botica-artalejo_rapid_2026}, we put forth the idea of "Nuclear Materials for AI" rather than the more frequently requested "AI for Nuclear Materials". This approach invites laboratories to re-frame their unit process of inquiry from absolute to relative, from single-material-at-high-quality to good-enough-is-good-enough, from seeking physics-first perfection first to generating enough data to train AI models of materials discovery, which can then be used to identify human-recognizable patterns which lead to key physical insights.

\section{Conclusion}
A high-throughput experimental framework for the accelerated down-selection of fusion-relevant materials has been developed and tested. Using the selection of a material for RF antennas for fusion reactors as showcase, the Cu-Cr-Ta alloy system was explored. This compositional space was selected due to its favorable low-activation characteristics and the potential for improved mechanical performance through the formation of strengthening C15 Laves secondary phases associated with Cr$_2$Ta precipitates. By integrating PVD combinatorial synthesis, rapid irradiation screening, and multi-criteria materials evaluation, the present methodology enables efficient identification of optimized alloy compositions within complex ternary systems.

The developed workflow substantially reduces the experimental time traditionally associated with radiation-tolerant materials development. Conventional irradiation qualification approaches often require multiple independent irradiations, extended exposure times to achieve defect accumulation detectable by TEM, and time-intensive post-irradiation characterization and sample preparation, resulting in experimental timelines spanning several months. In contrast, the present approach enabled the screening of 76 unique Cu-Cr-Ta compositions in approximately 20 h, speeding up screening and selection times by more than 100 times. This significant increase in experimental throughput demonstrates the potential of combining combinatorial materials synthesis with \emph{in~situ} radiation damage diagnostics as a scalable strategy for accelerated materials discovery in extreme environments.

A systematic multi-criteria down-selection methodology was subsequently implemented to identify the most promising alloy composition for fusion RF antenna applications. Initial screening prioritized alloys with waste disposal ratings (WDR) below 1, ensuring classification as low-level waste (LLW), together with Cr/Ta ratios near 2 to promote C15 Laves phase formation while maintaining a balance between mechanical strengthening and electrical conductivity. The shortlisted alloys were further ranked according to radiation damage resistance, thermal diffusivity, electrical conductivity, and compositional criteria. This cumulative evaluation identified Cu-4.8 at.\%Cr-2.2 at.\%Ta as the most promising candidate composition for RF antennas for fusion reactors. CALPHAD calculations further predict microstructural stability up to approximately $1084^\circ$C, satisfying the anticipated operating conditions for the RF antenna systems considered in this work.

Overall, the methodology presented here establishes a versatile platform for rapid irradiation materials screening and optimization, with broader applicability to the accelerated development of structural and functional materials for fusion and other extreme-environment technologies.

\section*{Data availability statement}
The data and code that support the findings of this study are openly available in the following repository: https://github.com/elenabotica/HighThroughput-Materials-Downselection.git.

\section*{Declaration of generative AI and AI-assisted technologies in the manuscript preparation process}
During the initial preparation of this work, ChatGPT-5 model were used by the main author for English correction and expression. After using this tool, all authors reviewed and edited the content as needed, and take full responsibility for the content of the publication.  

\section*{CRedit authorship contribution statement}
\textbf{Elena Botica-Artalejo} Conceptualization, Data curation, Methodology, Formal Analysis, Investigation, Writing - Original Draft, Writing - Review \& Editing, Visualization.  \textbf{Josh Randolph} Methodology, Investigation, Writing - Review \& Editing. \textbf{Kevin B. Woller} Methodology, Investigation, Resources, Writing - Review \& Editing. \textbf{Gregory M. Wallace} Conceptualization, Investigation, Methodology, Supervision, Funding acquisition, Writing - Review \& Editing. \textbf{Michael P. Short} Conceptualization, Investigation, Resources, Methodology, Supervision, Writing - Review \& Editing.

\section*{Declaration of Competing Interest} 
The authors declare that they have no known competing financial interests or personal relationships that could have appeared to influence the work reported in this paper.

\section*{Funding sources}
This work was supported by the U.S. Department of Energy, Office of Science, Office of Fusion Energy Sciences under Award Number DE-SC0024307.

\section*{Acknowledgment}
This work was performed in part in the MIT.nano Characterization Facilities.

\printbibliography


\clearpage
\appendix

\setcounter{section}{0}
\setcounter{figure}{0}
\setcounter{table}{0}
\setcounter{equation}{0}
\setcounter{page}{1}

\renewcommand{\thesection}{S\arabic{section}}
\renewcommand{\thefigure}{S\arabic{figure}}
\renewcommand{\thetable}{S\arabic{table}}
\renewcommand{\theequation}{S\arabic{equation}}
\renewcommand{\thepage}{S\arabic{page}}

\renewcommand{\theHsection}{supp.\arabic{section}}
\renewcommand{\theHfigure}{supp.\arabic{figure}}
\renewcommand{\theHtable}{supp.\arabic{table}}
\renewcommand{\theHequation}{supp.\arabic{equation}}

\section*{Supplementary Material}

\input{Supplementary_material}

\end{document}

%% file: Supplementary_material.tex
\section{TGS setup parameters}
\begin{table}[h]
    \caption{Detailed summary of TGS system parameters. Spot sizes reported as $2\sigma$ Gaussian widths.\\}   
    \centering
    \begin{tabular}{c c}
    \toprule
        \textbf{Parameter} & \textbf{Value} \\
        \hline  
        Pump wavelength & 532 nm \\
        Pump energy per pulse & 10 $\mu$J \\
        Pump pulse duration & $<$500 ps \\
        Pump spot size (on-sample diameter) & 150 $\mu$m \\
        Pump firing frequency & 1 kHz \\
        Pump on-sample power & 7.36 mW \\
        Probe wavelength & 671 nm \\
        Probe frequency & Continuous Wave \\
        Probe spot size (on-sample diameter)& 90 $\mu$m \\
        Probe on-sample power & 4.6 mW \\
        Detector bandwidth & 50 kHz - 1 GHz (3 dB) \\
        Oscilloscope bandwidth & 5 GHz \\
        Averaged traces per second & 2000 \\
        Calibrated on-sample grating spacing & 3.5264 $\mu m$ \\
        Sample roughness& $R_a = 10$ \AA \\
        \hline
    \end{tabular}
    \label{tab:tgs_param}
\end{table}

\section{Analysis procedure for SAW frequency drop calculation}
\begin{figure}[h]
\centering
\includegraphics[width=0.9\textwidth]{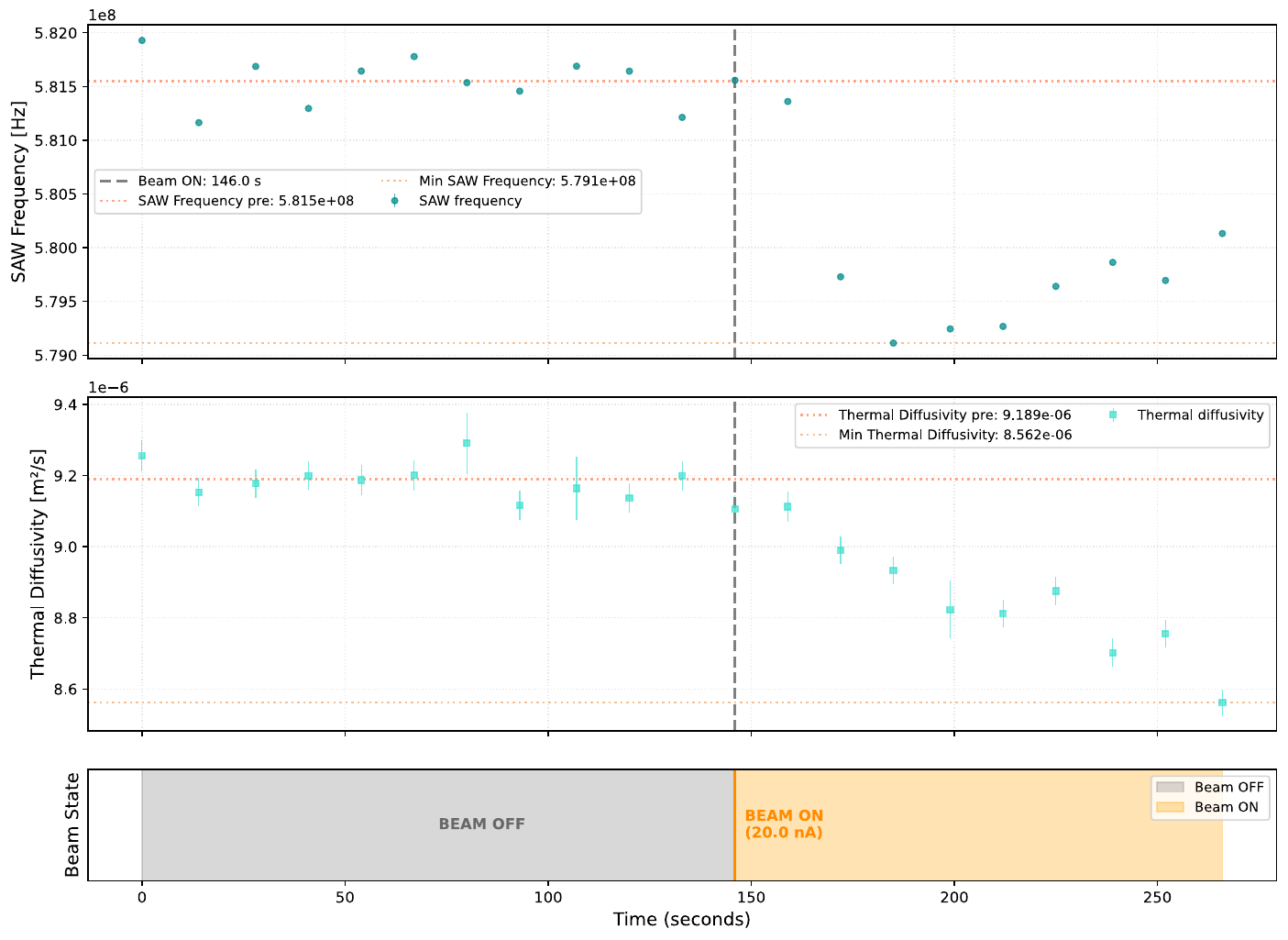}
\caption{Representative analysis of the I$^3$TGS signal for the Cu–5.2 at.\%Cr–2.4 at.\%Ta composition. The upper panel shows the evolution of the SAW frequency during ion irradiation, the middle panel shows the corresponding thermal diffusivity evolution, and the lower panel indicates the ion beam state and current.}
\label{fig:I3TGS_beam}
\end{figure}

In the analysis procedure, the initial SAW frequency was defined as the average frequency measured prior to beam exposure, while the minimum frequency reached during irradiation was identified automatically. The relative SAW frequency drop was subsequently calculated from these two values. An example of the signal processing methodology is shown in Figure \ref{fig:I3TGS_beam}.

\section{CALPHAD phase stability calculation}
\begin{figure}[h]
\centering
\includegraphics[width=\textwidth]{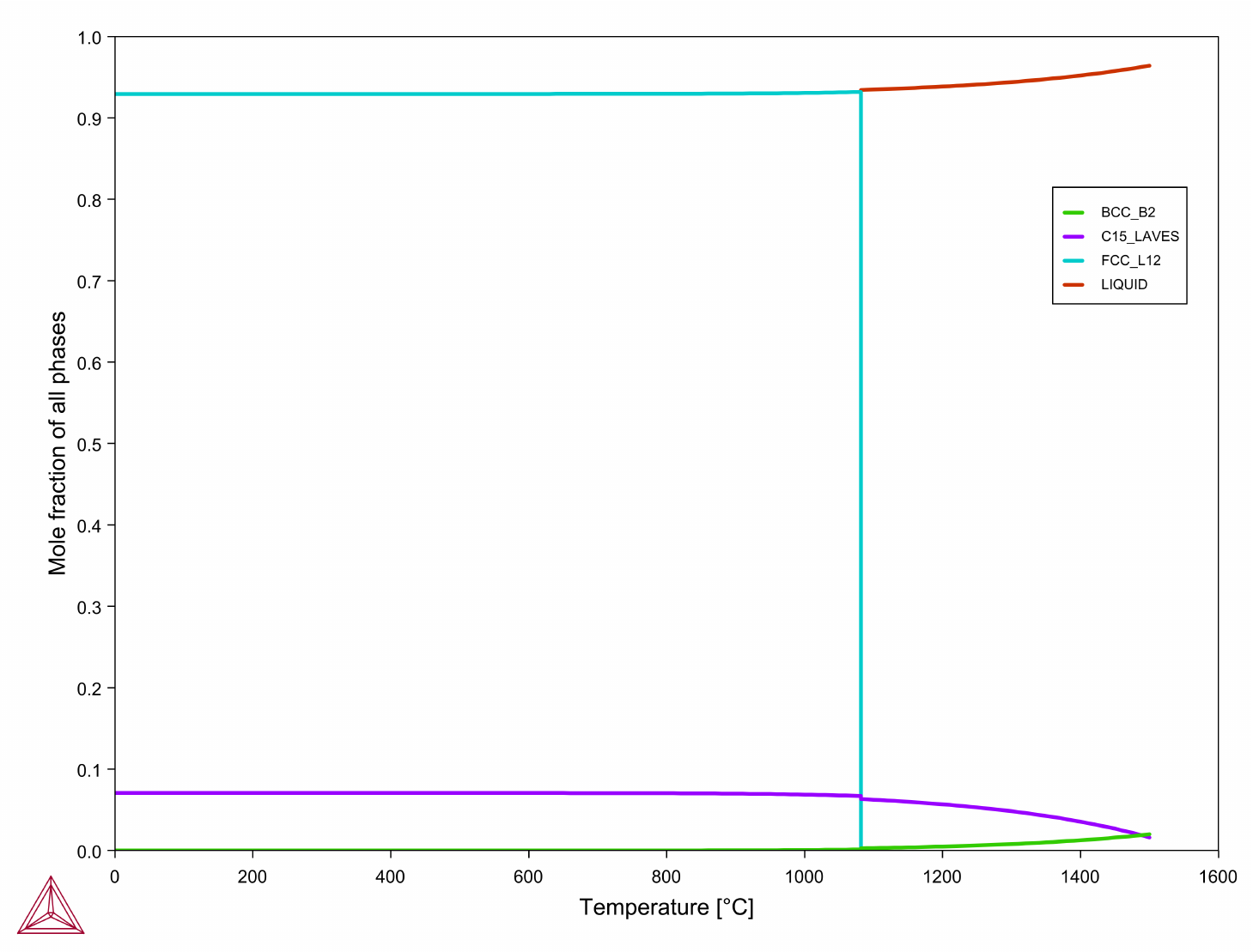}
\caption{CALPHAD-calculated phase fractions as a function of temperature for the Cu–4.8 at.\%Cr–2.2 at.\%Ta alloy using the TCHEA8 database in Thermo-Calc 2025b.}
\label{fig:TC_phase}
\end{figure}

To assure that the most promising material from the analysis is stable at the operating temperature for this type of RF antennas, CALPHAD calculations for phase stability were performed using Thermo-Calc 2025b in combination with the TCHEA8 database, see Figure \ref{fig:TC_phase}.

